\RequirePackage{lineno} 
\documentclass[twocolumn,twoside,slac_two]{revtex4}
\usepackage{graphicx}
\usepackage{fancyhdr}
\usepackage{graphics}
\usepackage{epstopdf}
\usepackage{textpos}
\newcommand{\EPC}[3]  {Eur.\ Phys.\ J.\ \textbf{C #1} (#2) #3}

\newcommand{\lnu}{\ensuremath{\ell\nu_\ell}}
\newcommand{\mtw}{\ensuremath{m_{\mathrm{T}}(W)}}
\newcommand{\sigmattbar}{$\sigma_{\ttbar}$}

\newcommand{\pp}{$pp$}
\newcommand{\intlum}{$\int${\it L}dt}
\def\antibar#1{\ensuremath{#1\bar{#1}}}

\def\ttbar{\antibar{t}}
\def\ppbar{\antibar{p}}
\def\roots {\ensuremath{\sqrt{s}}}

\def\ifb{\mbox{fb$^{-1}$}}
\def\ipb{\mbox{pb$^{-1}$}}

\def\pt{\ensuremath{p_{\mathrm{T}}}} 
\def\HT{\ensuremath{H_{\mathrm{T}}}} 
\def\ETmiss{\ensuremath{E_{\mathrm{T}}^{miss}}} 

\def\Ztau{\ensuremath{Z \rightarrow \tau\tau}}
\def\Wboson{\ensuremath{W}}%
\def\Zboson{\ensuremath{Z}}

\def\etal{\mbox{{\it et al.}}}
\def\atlasgaad{ATLAS Collaboration, G.\ Aad \etal}
\def\cmschat{CMS Collaboration, S.\ Chatrchyan \etal}

\begin{document}
\title{\centering Top Quark  Production at the LHC}
\author{
\centering
\begin{center}
Francesco Span\`o
\end{center}}
\affiliation{\centering
 Royal Holloway The University of London - Egham Surrey TW20 0EX - United Kingdom,
On behalf of the ATLAS and CMS collaborations
}    
\begin{abstract}
Top quark production in proton proton collisions at the Large Hadron
Collider (LHC) is reviewed using data collected by the ATLAS and CMS detectors.
Most recent results on searches for new physics related  to top quark
production mechanism are included.
\end{abstract}

\maketitle
\thispagestyle{fancy}

\section{Introduction}
The top quark is the most massive known fundamental constituent of
nature, the only one with a mass of the same order magnitude of the electroweak
symmetry scale. This property hints at an important connection with the
still mysterious mechanism of spontaneous symmetry breaking. 
While the study of its production and decay allow for a precision test of the standard
model (SM) predictions, the top quark is a ubiquitous ingredient for scenarios featuring
physics beyond our present understanding~\cite{theubiquitoustop}.
In models where spontaneous symmetry breaking results from the
presence of the Higgs boson or its supersymmetric extensions,  the top
quark is a crucial ingredient to constrain the Higgs mass
range~\cite{topForHiggs} and it is an important background to the
corresponding searches. In alternative scenarios featuring extra
dimensions or new strong forces, the top quark is often the preferred
object new physics couples to so as to modify its production and/or its decay
mechanism with respect to the SM predictions.

\vspace{-0.3 cm}
\section{ The Large Hadron Collider: producer of top quarks}
Copiuos top quarks production is desirable to perform a detailed
study of their properties. The proton-proton (\pp) collisions  with a center
of mass (\roots) of 7 TeV realized at the Large Hadron Collider
(LHC)~\cite{LHC} allow to explore top quark production at
unprecedented energy densities and  abundance.

In summer 2011 the LHC has already achieved its performance goals for
the year by reaching a peak luminosity {\it L} = $ 2\cdot
10^{33} \mathrm{cm}^{-2} \mathrm{s}^{-1}$, about ten times the one attained in
2010 ( $2.1 \cdot 10^{32} \mathrm{cm}^{-2} \mathrm{s}^{-1}$) and by
delivering an integrated luminosity per
experiment \intlum = 2.5 \ifb, fifty times larger than the one delivered
in 2010 (50 \ipb).

While the available luminosity encapsulates the space-time density of
the LHC collisions,  the other factor determining the number of events
with top quarks produced at LHC is the cross section for producing top
quarks in \pp\ collisions and its dependence on \roots.
At the LHC top quarks are predominantly produced in top/anti-top quark
pairs (\ttbar).  
The \ttbar\ production cross section, $\sigma_{\ttbar}$, is dominated by
the gluon fusion process over quark annihilation due to the relative
size of the gluon and quark parton distribution functions in the low proton
momentum fraction  region ({\em x} $\approx$ 0.025) probed by \ttbar\
events at the LHC~\cite{topPDFAtLHC}. 
The gluon fusion process accounts for about 85\% of $\sigma_{\ttbar}$
in \pp\ collisions at \roots\ = 7 TeV (and ~90\% at \roots\ = 10 TeV), thus
inverting the hierarchy observed in \ppbar\ collisions at
Tevatron~\cite{topXsecAtTev} ({\em x} $\approx$ 0.2 for \ttbar\ events).
The value of $\sigma_{\ttbar}$ at LHC with \roots\ = 7 TeV is estimated to be
165$^{+11}_{-16}$ pb at approximate next-to-next-to-leading order
(NNLO)~\cite{topPairXsecAtLHC}.
The electroweak single top (or anti-top) production cross section is about one third
of $\sigma_{\ttbar}$. It is characterized by the \Wboson\ boson mediated
$t$-($\sigma_{t}$ = 64 $\pm$ 3 pb~\cite{singleTopLHCTchan})  and
$s$-($\sigma_{s}$ = 4.6 $\pm$ 0.3~\cite{singleTopLHCSchan}) channel
production accompanied by the {\em Wt}-channel where a virtual $b$-quark mediates the associated
production of a \Wboson\ boson and a top quark in the final state
($\sigma_{Wt}$ = $15.7^{+1.3}_{-1.4}$pb~\cite{singleTopLHCWtChan}).

The cross section for massive particles' production increases with a
power law as a function of \roots\ in \ppbar/\pp\
collisions~\cite{CataniOnQCD}.
At LHC  with \roots\ = 7 TeV $\sigma_{\ttbar}$ increases by about a
factor twenty-three with respect to Tevatron. At the summer 2011 LHC
luminosity, this results in the production of about one \ttbar\ event every two seconds.
With $\int$ Ldt = 1\ifb\ the LHC expects 
about 2.4 times as many \ttbar\ events as those available in Tevatron
data with ninefold larger integrated luminosity.
\vspace{-0.3 cm}
\section{ATLAS and CMS: observers of top quarks}

As the top quark decays to a \Wboson\ boson and a $b$-quark about 100\% of the times, 
 the final state of a \ttbar\ event is characterized by the number of \Wboson\ bosons decaying
to a lepton-neutrino pair~\footnote{The W $\rightarrow$ \lnu\
  (q$\overline{q'}$) decay occurs  32.4\%  (67.6\%) of the times where
  q is a light quark, $\ell$ is a  lepton, $\nu_{\ell}$ is the corresponding lepton neutrino.}. The
case when both \Wboson\ bosons decay hadronically represents 45.7\% of the
events (fully hadronic channel),  while 34.3\% of the events feature
only one \Wboson\ boson decaying to a lepton (electron ($e$), muon
($\mu$) or tau ($\tau$) decaying to leptons) and  a neutrino
($\nu_{\ell}$) (single lepton channel).  The leptonic decay of both \Wboson\
bosons to $e$, $\mu$  or $\tau \rightarrow$ leptons occurs 6.5\% of the
time with the remaining 13.5\%  corresponding to double hadronic
$\tau$ decays.
The \ttbar\ final state features $b$-jets from $b$-quark production, high \pt\
jets from hadronic \Wboson\ boson decays,  at least one or two high $p_{t}$
leptons and large missing transverse energy
(\ETmiss) due to the neutrino in the \Wboson\ boson leptonic decays.
The final states of single top quark and~\ttbar\ can be
obtained from one another by swapping one $t \rightarrow Wb$ leg of
the~\ttbar\ decay with a \Wboson\ boson ($Wt$ channel) or
one/two quarks ($s$ and $t$-channels), one of which is a $b$-quark.
The two final states then have similar backgrounds (single
bosons (\Wboson, \Zboson) plus jets, di-bosons and Quantum Chromodynamics (QCD)
multi-jet events) and they are background to each other.

Such complex final states require the full involvement of the two
complementary, multi purpose detectors aimed at measuring the properties of leptons,
hadrons and photons in the LHC \pp\ collisions: ATLAS~\cite{atlasDet}
and CMS~\cite{cmsDet}. They feature layers of sub-detectors radially
expanding outwards with cylindrical symmetry around the line of the
colliding proton beams. From the tracking devices to the electromagnetic and hadronic
calorimeters to the muon stations, featuring different bending
magnetic field (solenoidal in CMS, solenoidal and toroidal in ATLAS) each entire
detector is at play in reconstructing  events with top quarks efficienty
selected by its three-(ATLAS) or two-(CMS) tier trigger system.
The excellent data-taking performance of the two detectors allows
each collaboration to analyze  \intlum\ = 36 \ipb\ of data collected
in 2010 and already \intlum\ = 0.2 to 1.4 \ifb\  of the \intlum\ = 2.5
\ifb\ of data recorded  per experiment up to summer 2011~\footnote{As
  these proceedings  are being  written ATLAS and CMS have completed
  their  2011 data collection for  \pp\ collisions with \intlum\
  $\approx$ 5.2 \ifb\ recorded per
  experiment.}. The values of \intlum\ are known at the level of 3.4\%
to 4.5\% .
\vspace{-0.3 cm}
\section{Ingredients for top quark detection with ATLAS and CMS}

The multiple ingredients to reconstruct the final state of events with top
quarks are the starting points of any top analysis.

Electrons~\cite{ATLASEmScale,CMSEmScale} are
defined as  isolated central objects ($|\eta_{e}|<2.4 (2.5)$ for ATLAS
(CMS)) with large transverse momentum (\pt) ($p_{T}^{e} >25$ (30) GeV
  for ATLAS (CMS) ) combining the electromagnetic shower
shape information of clusters in the calorimetry with the space-matched
tracks reconstructed by the tracking system.  ATLAS features an electron energy scale known within 0.3\% to 1.6\%  (up to 1 TeV) ~\cite{ATLASEmScale}  while CMS ECAL scale is
 known within 0.6\% to 1.5\%~\cite{CMSEmScale}. 
Duplicate electrons are removed either as reconstructed objects
recognized by the particle flow scheme used by CMS~\cite{CMSParticleFlow} or
(similarly) as close-by jets with $\Delta$R($e$, jet)$<$0.2~\footnote{The $\Delta$R distance
  between two particles with four-momenta $i$ and $j$ is defined as $\sqrt{(\phi_{i}-\phi_{j})^2
    +(\eta_{i}-\eta_{j}) ^{2}}$ where $\phi_{i}$, $\eta_{i}$ ($\phi_{j}$, $\eta_{j}$) are respectively
  the azimuthal angle and the pseudorapidity of particle $i$ ($j$).} rejected by
ATLAS.
\newline \indent
Muons~\cite{MuonsInAtlas, MuonsInCMS} are isolated
central ($|\eta_{\mu}|<2.5$ (2.1) in ATLAS (CMS)), high
$p_{T}$ (\ensuremath{p_{\mathrm{T}}^{\mu}}$>$ 20 GeV) tracks obtained from a combined fit of
information from the tracker and the external muon system.  The $p_{T}$
scale is known at about 1\% level.
An event is rejected when $\Delta$R($\mu$,jet)$ <$
0.4 (for ATLAS)  or 0.3 (for CMS) for at least one jet, to help
 suppress contributions from non-W-boson derived
muons (from flavour decays).
\newline \indent
Jets~\cite{jetsInATLAS,jetsInCMS} are reconstructed by feeding particle
flow objects (CMS) or calorimetric, three-dimensional,
noise-suppressed clusters (ATLAS)  to the anti-$k_{T}$ algorithm. 
Calibrated jets are obtained with an ($\eta$, \pt) dependent
weight from  simulated ``true'' kinematic information. 
The resulting jets need to be central ($|\eta_{jet}| < 2.5 (2.4)$ for ATLAS
(CMS)) with high \pt\ (\ensuremath{p_{\mathrm{T,jet}}} $>$ 25 (30) GeV for ATLAS (CMS)).
The energy scale uncertainty for jets ranges between $\approx$ 2\% and
8\%  as a
function of $\eta$ and \pt~\footnote{The typical expected inclusive central jet \pt\ range for
  selected \ttbar\ single lepton events at LHC with \roots\ = 7 TeV is
  between 25 or 30 GeV  (uncertainty $\approx$  4 to 8\%) and
  O(250) GeV with a mean \pt\ of $\approx$  70 GeV
  (uncertainty $\approx$ 2\%).}. The contributions
to the uncertainty include physics modelling, calorimeter response and detector simulation.
\newline \indent
Missing transverse energy is derived from the negative vector sum of
four momenta  of the objects in the event.
In ATLAS~\cite{etMissInATLAS} energy in calorimeter cells associated
with high \pt\ objects is summed with the muon momentum and an
estimate of the dead material loss. In CMS~\cite{etMissInCMS}  more than one technique is
used starting from calorimetric energy /momentum information, then adding track
information to it and/or finally considering the full set of objects
reconstructed with a particle flow technique. All the elements are
calibrated according to the high \pt\ object they are associated with.


\vspace{-0.3 cm}
\section{\ttbar\ production: single lepton channel}
\label{sec:singleLepNoBtag}

Both ATLAS~\cite{ATLASSingleLep} and CMS~\cite{CMSSingleLep}  measured
\sigmattbar\ using single lepton events.
The final state is characterized by one high \pt, central
lepton ($e$ or $\mu$) with at least
three jets resulting from the hadronic decay of the top quark. 
While in ATLAS large \ETmiss\ and transverse mass of the leptonic \Wboson\ boson (\mtw)  are required to reduce the impact of
QCD, in CMS these cuts are not applied as \ETmiss\ is used as a variable in a likelihood fit.
The basic cuts select $\approx$ ten to twenty  (one) thousand events for
\intlum\ = 0.7\ifb (36 \ipb).
\newline \indent
The dominant backgrounds are \Wboson\ + jets events and multi-jet
events from QCD.
Both these backgrounds are constrained from data.
The \Wboson\ + jets shape is
derived from simulation; its normalization is left as a
parameter for a likelihood fit to determine.
 (ATLAS sets its initial value and Gaussian constraints for the fit
 from the asymmetry in \Wboson\ boson
production in \pp\ collisions.) 
ATLAS derives the shape, the initial value for the QCD background
normalization  by combining the content of QCD-enriched
control samples derived from non-isolated
leptons with the probabiities that real lepton and fake leptons meet
isolation requirements. CMS considers control samples based on events failing only the electron identification
requirements. 
The final QCD normalization is a floating parameter in the likelihood
fit (with data-driven costraint for ATLAS).
The shape of smaller electroweak backgrounds from single top events,
di-boson production and Z/$\gamma$ + jets is derived from simulation
and the normalization is a fit-determined parameter.
\newline \indent
In both analyses a discriminant is the built from signal and background templates of kinematic quantities.
ATLAS uses lepton $\eta$, the \pt\ of
the leading jet and variable related to how spherical and how
transverse the event is (aplanarity).
 CMS uses \ETmiss\ for
  events with three jets  
and the mass of the three-jet system
with the highest vectorially combined \pt\ for the events with four
or more jets. A binned maximum likelihood fit of the discriminant templates to the
data is performed to extract the cross section for the signal and the
backgrounds in either three-, four- and more than four-jet samples
(ATLAS) or only in the three- and four-or-more-jet samples (CMS).
\newline \indent
For ATLAS the likelihood fit includes systematic uncertainties as
nuisance parameters to be constrained from data, thus resulting into a
reduction from 20\% to 70 \%  of their contribution. The ATLAS  result using
\intlum\ $\approx$ 0.7 \ifb\ is $\sigma_{\ttbar}$= 179.0 $\pm $3.9 (stat.)
$\pm$ 9.0 (syst.) $\pm$ 6.6 (lumi.) pb with a total relative uncertainty
of 6.6\% where about 5\% (in quadrature) is of systematic origin.
CMS includes systematic uncertainties
  in the pseudo-experiment used to derive the Neyman confidence level (CL)
  belt.
  The CMS result for \intlum = 36 \ipb\ is
  $\sigma_{\ttbar}$= 173 $\pm$ 14 (stat.)$^{\pm36 }_{\pm 29}$ $\pm$ 7
  (lumi.) pb and it is dominated by systematic uncertainties
  (mostly jet energy scale) accounting for with 21\%  of the total 23\% uncertainty.
\vspace{-0.2cm}
\section{Ingredients for top quark detection: enter $b$-jets}

The remaining ingredient used in characterizing the top quark final state,
$b$-jets, uses the fact that $b$-hadrons are characterized by a non-zero
observable flight distance from the primary vertex and the presence of
tracks with non-zero distance of closest approach with respect to the
primary vertex. These properties together with the number of tracks
related to the secondary vertex and their energies are the basis for a
series of discriminants that both ATLAS~\cite{bjetInATLAS} and
CMS~\cite{bjetInCMS} use to separate $b$-jets from other types of
jets. 
In both cases  the performance of the $b$-jet
identification ($b$-tagging) is assessed by using $b$-jet enriched control samples
while the rate of mis-tagging is obtained by events
characterized by the negative values of secondary vertex properties. 
Large efficiencies at the level of 80\% are coupled to mis-tagging
probabilities around 10\%, while lower efficiencies around 40\% allow
purer samples with only a 0.1\% mis-tagging rate. 
\vspace{-0.3 cm}
\section{\ttbar\ production: single lepton channel with $b$-tagging}
\label{sec:singleLeptBtag}
The standard single lepton selection and
background assessment outlined in section~\ref{sec:singleLepNoBtag}
are complemented by requiring at least one $b$-tagged central, high \pt\ jet.
CMS~\cite{CMSSingleLepBtag}  performs a
maximum likelihood fit to the secondary vertex mass in the two
dimensional plane of standard and $b$-tagged jet
multiplicity.
 ATLAS~\cite{ATLASSingleLepBtag} uses the same maximum
likelihood fit to a four-variable discriminant used in~\cite{ATLASSingleLep}, but
it replaces the leading jet \pt\ with the average of the two largest
jet $b$-tagging probabilities.  In both cases systematic uncertainties
are extracted from the fit as nuisance parameters.
Both analyses use \intlum\ = 36 \ipb.
 The results are  $\sigma_{\ttbar}$= 150 $\pm$ 9 (stat.) $\pm$ 17
(syst.) $\pm$ 6 (lumi.) pb for CMS and $\sigma_{\ttbar}$= 186 $\pm$ 10
(stat) $^{+ 21}_{-20} $ (syst.) $\pm$ 6 (lumi.) pb for ATLAS. Both
results have the same relative uncertainty of about 13\%.
\vspace{-0.3 cm}
\section{\ttbar\ production: di-lepton channel}
\label{sec:diLept}
The di-lepton final state is analyzed by both
ATLAS~\cite{ATLASDiLepton} and CMS~\cite{CMSDiLepton}. After
requiring a single lepton (ATLAS) or even a di-electron (CMS)
trigger, the di-lepton final state is characterized by exactly (ATLAS) or at
least (CMS) two opposite-sign high \pt\ central leptons ($e$ or
$\mu$), at least two central, high \pt\ jets (from $b$-quarks) and large \ETmiss 
(in the di-electron or di-muon case) or large transverse activity (\HT\  = sum of $|$\pt$|$ of jets and leptons  (ATLAS) or sum of
leptons' transverse masses (CMS)). 
 In addition events with  di-lepton masses that are either \Zboson\
 boson-like or below 15 GeV are vetoed as the backgrounds are the same as the single lepton channel with the
replacement of \Wboson\ + jets with \Zboson/$\gamma$ + jets. In case at least one $b$-tagged jet
is required the \ETmiss\ requirement is relaxed. 
\newline \indent
The fake lepton background deriving from QCD is estimated from data.
 In a generalization of the technique used for single-lepton analyses (see
section~\ref{sec:singleLepNoBtag}),  the probability that either a
loosely selected fake or a real lepton is selected in the signal region
is derived in control samples enriched with real or fake
leptons (ATLAS) (\Zboson-like and low \ETmiss\ events
respectively) or in a multi-jet enriched single loose-lepton
sample (CMS).
These probabilities are then combined with the number of di-lepton
events featuring either one of the three combination of
(tightly, loosely) selected lepton pairs (ATLAS) or only
a pair of loosely selected leptons (CMS): the
number of fake leptons in the signal region ( i.e. tight leptons) is the result.
\newline \indent
In parallel the \Zboson/$\gamma$ + jets background is obtained by subtracting the
expected non-\Zboson/$\gamma$+ jets simulated-background from the  observed number of events in
the \Zboson-mass window control region and scaling the resulting value with
the simulated ratio of the expected number of Z/$\gamma$ + jets events in the
control region to the the number of the same events expected in the signal region.
The remaining electroweak backgrounds are estimated from simulation.
The expected signal to background ratio ranges from 8 to 10 with a good
agreement between data and simulation for about 2500
(1100) events in \intlum\ = 1.1
\ifb for CMS (0.7\ifb\ for ATLAS).
\newline \indent
The value of \sigmattbar\ is extracted by a maximum likelihood fit
to a counting experiment hypothesis only incorporating the number of
signal events expected in the three channels, including the estimates
of the backgrounds and adding systematic uncertainties as nuisance
parameters in the fit.
The results obtained by ATLAS  using \intlum\ $\approx$ 0.7 \ifb\ with
and without the requirement of at least one $b$-tagged jet are:
\sigmattbar\ = 171 $\pm$ 6 (stat.) $^{+16}_{-14}$ (syst.) $\pm$ 8 (lumi.) pb  (untagged) and
\sigmattbar\ = 177 $\pm$ 6 $^{+17}_{-14}$ (syst.)  $^{+8}_{-7}$ (lumi.) pb ( $\geq$1 $b$-tag).
CMS untagged result  with \intlum=1.14 \ifb\ is $\sigma_{tt}$ = 169.9
$\pm$ 3.9 (stat.) $\pm$ 16.3 syst.) $\pm$ 7.6 (lumi.).
All results have a relative uncertainty around 11\% that is already
systematics-dominated. For ATLAS  jet energy scale (~5\%) or $b$-tagging
(5\%) are dominant, while for CMS pile-up and lepton selection account
for about 5\% of the total uncertainty.
\newline
\subsection{Using taus: the $\tau \mu$ channel}
\label{sec:diLeptTauMu}
Both ATLAS~\cite{ATLASTauMu} and CMS~\cite{CMSTauMu} have started
using hadronic tau ($\tau$) decays in the di-lepton final states.
A high \pt\ muon is requested to be
accompanied by at least one jet-seeded $\tau$ candidate with opposite
sign to the muon, either resulting from a cut-flow-based algorithm
run on particle flow objects (CMS) or detected by a boosted decision
tree scheme (ATLAS).
Then (at least) two jets and
at least one $b$-tagged jet are required together with large \ETmiss and
large \HT. The dominant backgrounds (\ttbar\ and
\Wboson\ + jets) are derived from data either in the low jet multiplicity
region (ATLAS) or by weighting a \Wboson+$\geq$3 jets enriched sample
with the $\tau$-faking probability derived from averaging two data-driven
estimates in \Wboson\ +1 jet sample and QCD enriched regions
(CMS). The QCD shape is derived from a control region  with non
isolated muons and normalized to the low \ETmiss\ region.


The value of \sigmattbar\ is then obtained by scaling the number of signal
events with the acceptance and luminosity. The number of signal events
are obtained by either subtracting the estimated background (CMS) or 
by a likelihood fit to the distribution of the boosted decision tree
variable used for $\tau$-tagging (ATLAS). Such distribution is derived
by taking the difference between samples with opposite sign and same sign
$\tau$$\mu$ pairs (so as to cancel out the presence of most events
where a gluon- or $b$-jet fakes a $\tau$ candidate).


The resulting cross sections  using \intlum = 1.08 \ifb\ are
$\sigma_{tt}$ = 142 $\pm$ 21 (stat.) $^{+20}_{-16}$ (syst.) $\pm$ 5
(lumi.) pb (ATLAS) and $\sigma_{tt}$ = 148.7 $\pm$ 23.6 (stat.) $\pm$
26.0 (syst.) $\pm$ 8.9 (lumi.) pb. 
The results have comparable relative uncertainties of 24\%
(CMS) and 21\% (ATLAS)  with similar sizes for statistical
and systematic uncertainties.
\vspace{-0.3 cm}
\section{\ttbar\ production: fully hadronic channel}
\label{sec:fullyhad}
The \ttbar\ fully hadronic final state is  analyzed by both ATLAS~\cite{ATLASFullyHad} and
CMS~\cite{CMSFullyHad}. Events are selected by requiring a multi-jet (at
least 4 jets) trigger and at least six high \pt\ central jets with at
least two $b$-tagged jets. In ATLAS no electrons or muons are
allowed in the final state and small \ETmiss significance
(\ETmiss/\roots($E_{T,calo}$)) and large \HT\ are required.
Both analyses reconstruct the events with a least-squares ($\chi^{2}$) kinematic fit to
the \ttbar\ hypothesis.
The dominant QCD background is derived from data by weighting events
from an untagged multi-jet control sample with five or six jets with a
data-driven $b$-tagging probability.  The number of signal events is
extracted from a likelihood fit to either the top quark mass distribution
(also checked by a neural network discriminant) (CMS) or to the 
$\chi^{2}$ distribution from the fit (ATLAS). 
The value of \sigmattbar\ is obtained by scaling the number of signal
events with acceptance and luminosity.
Systematic uncertainties are dominated by contributions from $b$-tagging, jet
energy scale and background normalization. With \intlum\
=36 \ipb\ ATLAS sets a 95\% confidence level upper limit of
\sigmattbar\ $<$ 261 pb. Using \intlum= 1\ifb\ CMS obtains
\sigmattbar\ = 136 $\pm$ 20 (stat.) $\pm$ 40 (syst.) $\pm$ 8 (lumi.):
this is already systematics-dominated with a relative uncertainty of 33\%.

\vspace{-0.3 cm}
\section{Combined results for \ttbar\ production}
Both ATLAS~\cite{ATLASCombine} and CMS~\cite{CMSSingleLepBtag} combine their measurements of  \sigmattbar\  from
the single lepton and di-lepton channels as shown
in figure~\ref{ATLASAndCMSCombine}. The combined results for
\sigmattbar\ are:

\vspace{0.5mm}
$\mathrm{ATLAS}: 176 \pm 5 \mathrm{
  (stat.)}^{+13}_{-10}\mathrm{(syst.)} \pm 7 \mathrm{(lumi.)}
\mathrm{pb}$, %

\vspace{0.5mm}

$\mathrm{CMS}$:     $154 \pm 17 \mathrm{ (stat. + syst.)} \pm 6\mathrm{(lumi.)} \mathrm{pb}$.
\vspace{0.5mm}

\noindent These results do not include the latest untagged single-lepton measurement by
ATLAS using \intlum\ = 0.7 \ifb\ (see section~\ref{sec:singleLepNoBtag}) and the
latest di-lepton measurement by CMS using \intlum\ = 1.1.4 \ifb\ (see
section~\ref{sec:diLept}). The relative uncertainty is in both cases at the level of 10\%: it is comparable with the theoretical uncertainty
and already systematics-dominated.
\begin{figure}[hbt]
\includegraphics[width=65mm]{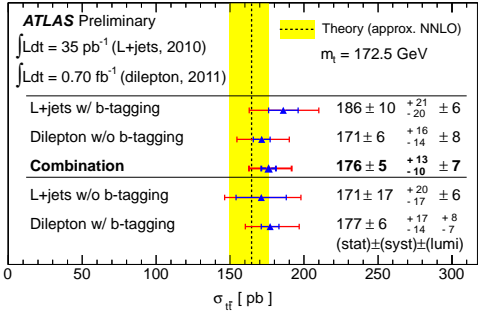}
\includegraphics[width=65mm]{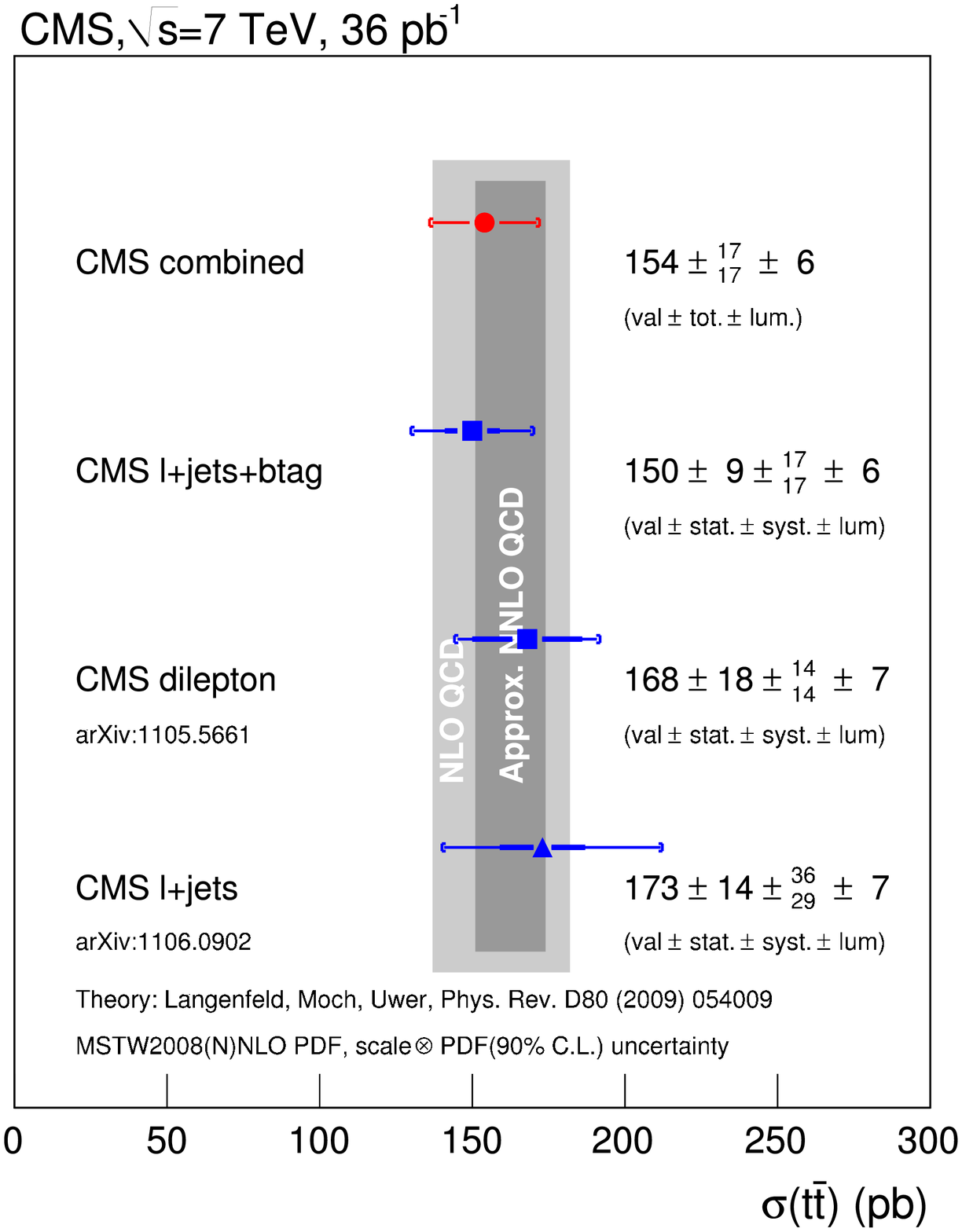}
\caption{Available summaries and combinations of \sigmattbar\ measurements from
  ATLAS~\cite{ATLASCombine} (upper figure) and
  CMS~\cite{CMSSingleLepBtag} (lower
  figure)\label{ATLASAndCMSCombine}. See text for comments on required updates.}
\end{figure}

\vspace{-0.3 cm}
\section{Single top quark production}

\subsection{The {\em t}-chnnel}
The single top {\em t}-channel events are selected by requiring
one central, high \pt\ lepton ($e$ or $\mu$), large \ETmiss\ and
 \mtw\ and exactly two or three jets with $|\eta_{jet}|<$ 4.5
(ATLAS) or 5 (CMS). Both ATLAS~\cite{ATLASSingleTopTChan} and
CMS~\cite{CMSSingleTopTChan} consider samples with and without
$b$-tagging requirements and derive QCD and \Wboson\ + jets normalization from data.
CMS combines two results: a two-dimensional maximum likelihood fit to
the data in the space of the angle between the lepton and the
untagged jet in the top-quark rest frame and the $\eta$ of the
untagged jet;  a Bayesian estimate of the cross section derived
from a boosted decision tree discriminant. 
ATLAS uses a cut-based analysis involving jet angular variables, \HT\
and the leptonic top quark  mass. The result is also confirmed by a maximum likelihood fit to a neural network output based on a set of thirteen
discriminating variables. ATLAS uses \intlum = 0.7 \ifb\ to obtain
$\sigma_{t}$ = $\sigma_{t}$= 90 $\pm$
9 (stat.) $^{+31} _{-20}$ (syst.) (with a significance  of 7.6
standard deviations (s.d.)), while CMS result with \intlum\ = 36
\ipb\ is 83.6 $\pm$ 29.8 (stat.+syst.) $\pm$ 3.3 (lumi.) pb (with 3.5
s.d. significance).

Both results are systematics-dominated with the same relative uncertainty of 36\%.

\subsection{The {\em Wt}-channel}
The {\em Wt}-channel single top events are searched for in the fully
leptonic configuration by ATLAS~\cite{ATLASWtChanSearch} using the
standard di-lepton selection and adding the requirement of exactly one central high \pt\
jet (from the $b$-quark of the top quark decay) and a cut on the azimuthal angle
between the lepton and \ETmiss\ to reject \Ztau\ decays. 
The main backgrounds are derived from data control samples: QCD
uses lepton isolation information as in
section~\ref{sec:singleLepNoBtag}. The Z/$\gamma$ + jets background
is extrapolated within the (\ETmiss, di-lepton mass) plane, the specific
Z$\rightarrow \tau\tau$ decay is extrapolated from the region with low
sum of the angles between the lepton and \ETmiss. Finally the dominant
\ttbar\ contribution is extrapolated from two-jet events.
A cut-and-count analysis then uses a maximum likelihood fit to combine
the channels by fitting systematic uncertainties as nuisance parameters.
The good agreement between data and simulation induces ATLAS to set an
observed (expected) 95\%  CL level upper limit on $\sigma_{Wt}$ of 39
(41) pb using \intlum\ = 0.7 \ifb.

\subsection{The {\em s}-channel}
The $s$-channel single top production is searched for by
ATLAS~\cite{ATLASSingleTopSChanSearch} by using the
single lepton selection modified by the requirement of having exactly
two high \pt\ central jets and using the same triangular cut on \mtw\
for $e$ and $\mu$ to reject QCD. Events with and without $b$-tagged
jets are considered, while the analysis result requires two $b$-tagged jets.
The estimate of QCD is derived from data by fitting the normalization
of  the \ETmiss\ shape derived in a sample enriched in electron-like
jets. The \Wboson\ + jets normalization is extrapolated from combining
information from the untagged sample and from the one and two-jet bins of
the tagged samples. A cut and count analysis combines channels with a
maximum likelihood where systematic uncertainties are constrained from data.
The good data-to-simulation agreement supports setting an observed
(expected) 95\% CL limit on $\sigma_{s}$ of 26.5 (20.5) pb with \intlum\ =  0.7 \ifb.

\vspace{-0.3 cm}
\section{Top production as a window on new physics}

A variety of alternative scenario can account for deviations from the SM in \ttbar\ production.
The presence of large mass resonances decaying preferentially to
\ttbar\ is a widely studied scenario~\cite{maltoni}. At the highest
 \ttbar\ masses ($M_{\ttbar}$) it produces a di-top-jet topology in
which the final products of each top quark decay are closely merged
into a single jet (boosted)~\cite{boostedtops}. 
The \ttbar\ production can also result from the presence of a heavy partner
of the top quark ($T$) which is pair-produced and decays to a top quark and a neutral stable
particle ($A_{0}$) representing a good dark matter candidate. The resulting
final state features a pair of top quarks with an increased amount of
missing transverse energy from the dark matter
candidates~\cite{theoryTTPlusETmiss}. In addition the detection of same
sign top quarks can signal flavour changing neutral currents (FCNC) proposed as a possible
explanation for the recently observed discrepancy between data and SM
predictions in the \ttbar\ forward-backward
asymmetry~\cite{theoryFCNCTopAsy,dMiet,yPet}.

\subsection{Search for FCNC-induced same-sign top pair production}
 
CMS searched for same sign top  quark pair
production~\cite{CMSSameSignSearch} by requesting two positive
isolated leptons, at least two high \pt\ jets and large \ETmiss. The dominant background consists of single
lepton \ttbar\ events with one fake lepton and it is derived from a data
control sample selected with loose electron isolation and
identification requirements. The selected events show no
excess over the background. A Bayesian technique is used to include
systematic uncertainties and set a 95\% credible interval  in the
production of same sign top pairs as a function of the mass of the
mediating $Z^{\prime}$ and its right-handed chiral coupling.
The preferred region for an FCNC-explanation of the \ttbar\
forward-backward asymmetry is excluded. 
In particular for a $Z^{\prime}$ mass
of 2 TeV a more stringent limit than the recent Tevatron one is set
on the strength of effective four-fermion contact interactions ($\frac{C_{RR}}{\Lambda}$
$<$2.7 $\mathrm{TeV}^{-2}$).

\subsection{Search for excess in \ttbar\ production with large \ETmiss}

ATLAS searched for an excess in \ttbar\ events with
large \ETmiss~\cite{ATLASTtbarPlusEtmissSearch} with \intlum\ = 1.08 \ifb. 
The standard single lepton selection was enriched by requiring very
large \ETmiss  ($>$100 GeV),  large \mtw\ and
vetoing events with $b$-tagged jets or an  additional low \pt\
lepton.  The dominant \Wboson\ + jets and \ttbar\ background is estimated
from data: the shape is derived from low jet multiplicity
events with $b$-tagging veto; the normalization results from the low \mtw\
region. QCD background estimated from data (like in section~\ref{sec:singleLepNoBtag}) is found to
be negligible. No excess is found in the selected data over the
expected background. The event yields are used to build a frequentist
statistic~\cite{cls} to set a 95\% CL limit  
for the cross section times branching for  the
$T\overline{T}\rightarrow \ttbar\ A_{0}A_{0}$ reaction and as a
function of  $A^{0}$ mass ($m_{A^{0}}$) and $T$ mass ($m_{T}$).
For $m_{A^{0}} <$ 140 GeV, scenarios with  340 GeV $<m_{T}< $ 380 GeV are excluded. 
In particular for $m_{A^{0}}$ smaller than 30 GeV  the $m_{A^{0}}$ = 410 GeV scenario is excluded at 95\% CL. 
 Finally ($\sigma\times$BR) = 1.1 pb is excluded at 95\%CL for
 ($m_{A^{0}}$, $m_{A^{0}}$) = (420 GeV, 10 GeV).


\subsection{Search for excess in \ttbar\ production versus  $M_{\ttbar}$}

Both ATLAS~\cite{ATLASResonanceSearchSingleLep} and
CMS~\cite{CMSResonanceSearchSingleLep} searched for excess resonant
\ttbar\ production in the single lepton channel.
The standard  single lepton selection is used in ATLAS with
at least four jets and one $b$-tagged jet in the final state. CMS developed
a boosted top quark selection in the single muon channel, requiring at
least  two jets with \pt\ $>$50 GeV for which the leading jet 
\pt\ is larger than 250 GeV. One non-isolated muon is
required with either a $\Delta$R  distance of 0.5 or a relative \pt\ of al least
15 GeV with respect to the spatially closest jet. Finally a large  value for the
sum the lepton \pt\ plus \ETmiss is required.
The main backgrounds are estimated from data. For QCD ATLAS derives its
shape from a jet-enriched sample normalized to the low \ETmiss\
region while CMS uses the events failing the two-dimensional muon selection
cuts. The \Wboson\ + jets normalization is extrapolated from small jet
multiplicity events. After reconstructing the leptonic \Wboson\ boson using
the \Wboson\ mass constraint with \ETmiss\ and the $\mu$ four
momentum, the \ttbar\ mass is calculated by either summing the leptonic
\Wboson\ boson to the four leading \pt\ jets (ATLAS)  or to the jets that are
consistent with a back-to-back boosted di-jet topology (CMS).

No excess is observed and 95\% Bayesian credible intervals are set for
$Z^{\prime}$ and Randall-Sundrun (RS) KK-gluon production including systematic
uncertainties by either integrated (CMS) or averaged (ATLAS) nuisance parameters. 
CMS uses \intlum\ = 1.14 \ifb\ to set upper observed (expected)
limits at 95\% probability on narrow $Z^{\prime}$
($\Gamma_{Z^{\prime}}/M_{Z^{\prime}}$  = 1\%) $\sigma \times BR$  at the
sub-pb level for $M_{Z^{\prime} }$ $>$ 1.3 TeV and at less than 0.2 pb for
$M_{Z^{\prime}}$ $>$ 2.3 TeV. No narrow resonance scenarios are
excluded, however if a less-narrow $Z^{\prime}$
($\Gamma_{Z'}/M_{Z'}$ = 3\%) is considered, CMS excludes scenarios with
805 GeV $<M_{Z'}<$ 935 GeV  and 960 GeV $<M_{Z'}<$ 1060 GeV.  
ATLAS also excludes scenarios with KK-gluon masses below 650 GeV
with 95\%  probability.
\newline \indent
A search for resonant \ttbar\ production is also carried out by
ATLAS~\cite{ATLASResonanceSearchDiLep} in the di-lepton (e,$\mu$)
channel with \intlum\ = 1.04 \ifb. The standard di-lepton selection is applied and data
driven estimates were derived for QCD and Z/$\gamma$ + jets
backgrounds from a control sample selected with an \ETmiss-dependent
\Zboson-window for the di-lepton mass. No excess is found in the sum
of $H_{T}$  and \ETmiss. 
A 95\% Bayesian credible interval is obtained for $\sigma\times BR$
for the production of RS KK-gluon including systematic uncertainties
as integrated nuisance parameters. The standard
scenario for RS KK-gluon production is excluded with
95\% probability for KK Gluon masses below 0.84 TeV.
\newline \indent
CMS  also developed a search for boosted resonances in \ttbar\
hadronic decays~\cite{CMSResonanceSearchFullyHad}  triggered on events
with at least one jet with very high \pt\ ($>$ 200 GeV).
Two configurations are analyzed for \intlum\ $\approx$ 0.9 \ifb.  
The ``1+1'' configuration  requires at least two Cambridge-Achen  (CA)
large cone ($\Delta$R = 0.8) jets with \pt $>$ 350 GeV and large
$\delta\phi$ separation and with both jets tagged
as {\em top}-jets. Top-jet tagging is based of the consistency of the jet
mass with the top quark mass, the presence of at least three sub-jets
in the last two step of the CA algorithm with a minimum mass of at
least 50 GeV for at least one sub-jet pair. The ``1+2'' configuration
requires at least three large cone CA jets featuring one leading
top-tagged jet with \pt $>$ 350 GeV and a second (third) ``pruned'' jet
( i.e. with less soft, wide angle clusters) with \pt $>$ 200 (30) GeV
and large $\Delta\phi$ from the leading \pt\ jet. The second jet is
requested to be recognized as a \Wboson-jet i.e. with a jet mass
consistent with the \Wboson\ mass, two sub-jets  and a maximum ratio of the
sub-jet mass to the jet mass of 0.4. The second and third jet are
required to form a jet with a mass consistent with the top mass.
The dominant QCD background is obtained by weighting a di-jet control
  sample featuring one top-tagged jet with  data-driven mis-tagging probability. The
  mis-tagging probability is obtained from the fraction of
  top/W-tagged probe jets in QCD enriched high \pt\ di- and tri-jet
  events with one anti-top-tagged jet (failing  a subset of top- or \Wboson\-tagging cuts). 
 The mass of the \ttbar\ system is obtained by summing the top-jets in
 the ``1+1'' configuration or the top-jet, the \Wboson-jet and the closest
 jet to the \Wboson-jet in the'' 1+2'' configuration. 
For the QCD background the untagged jet mass is flatly randomly chosen
in the (140 GeV, 250 GeV) interval to provide similarity to the jets
in the signal region.
 No excess is observed and a 95\% Bayesian credible interval is
 derived for $\sigma \times BR$ for the
 production of both $Z^{\prime}$ and RS-like KK-gluons, including
 systematic uncertainty as integrated nuisance parameters. Sub-pb
 limits are obtained on  $Z^{\prime}$  $\sigma \times BR$  and
 scenarios with KK-gluon with a mass between 1 and 1.5 TeV are
 excluded with 95\% probability.

\vspace{-0.3 cm}
\section{Conclusions}
Top quark production analysis  at LHC is in full swing thanks to the combined
performance of the collider and the associated detectors: a very rich program
is already underway.

The  value of \sigmattbar\ is measured in nearly all expected final
states. It is consistent with the SM at \roots\ = 7 TeV and most
measurements are systematics-dominated, entering the
realm of precision physics with $\Delta$\sigmattbar/\sigmattbar $\leq$
10\%.
Single top production is clearly observed in the {\em t-}channel while  more data is needed to observe it in the {\em Wt} and {\em s}-channel.

The rapidly increasing data-set and detector understanding is quickly opening unprecedented phase space for new physics searches linked to 
top quark production, ranging from resonances  to dark matter
candidates, whose mass sensitivity is by now breaking the 1 TeV barrier.


\end{document}

\bibitem{tom} T.~Junk, Nucl. Instrum. Meth. A {\bf 434}, 435 (1999)
%

\item{topPairXsecLHC}Aliev et al 2011, Beneke et al 2010, Langefeld Moch Uwer 2009, Moch,Uwer 2008
\item{singleTopLHCTchan}Kidonakis 2010 (various)
\item{singleTopLHCSchan}Kidonakis 2010 (various)
\item{singleTopLHCWtChan}Kidonakis 2010 (various)
\item{si}David W these proceedings and references therein
\item{singletopxsec}David W these proceedings and references therein
\item{singletopxsec}David W these proceedings and references therein

\item{elScaleAtlas}
\item{slScaleCMS} 
\item{muScaleAtlas} ATLAS-CONF-2011-032, arxiv:1107.4277
\item{muScaleCMS} ATLAS-CONF-2011-032, arxiv:1107.4277
\item{jetScale} ATLAS-CONF-2011-032, arxiv:1107.4277
\item{etMissAtlas} ATLAS-CONF-2011-080
\item{etMissCMS}  arxiv:1106.5048v1

\item{atlasdilepton} https://atlas.web.cern.ch/Atlas/GROUPS/PHYSICS/CONFNOTES/ATLAS-CONF-2011-100/
\item{cmsdilepton} https://twiki.cern.ch/twiki/bin/view/CMSPublic/PhysicsResultsTOP11005
\item{atlastaumu}
\item{cmstaumu}
\item{cmstauid}
\item{atlastauid}